\documentstyle[]{article}
\title{Literal Rippling of Spacetime}
\author{N. Redington\\Building 14S - 100\\Massachusetts Institute of Technology\\Cambridge, MA 02139, USA}
\date{}
\begin {document}
\maketitle
\begin{center}
\begin{it}
ABSTRACT:
\end{it} \\

\begin{it}
The metric perturbation tensor corresponding
to a transverse oscillation of spacetime is 
composed of products of cosines. When averaged over many wavelengths,
such a metric may look either Minkowskian or Euclidean at large scales,
depending on the amplitude and wavelength of the oscillation.

\end{it}
   
\end{center}

\newpage

A curved surface may be described in terms of its intrinsic geometry 
or in relation to a higher-dimensional space in which it is embedded. 
In the case of a four-dimensional hypersurface, an embedding space of 
ten dimensions is required. It is well known that solutions of Einstein's 
equation can be embedded in a space of ten dimensions; see for example 
[Rosen 1965],  [Dirac 1975] or [Kramer {\it et al.} 1980]\@. 

Consider a surface which, like that of the ocean, experiences oscillations 
transverse to itself into the embedding space. If the surface is 
four-dimensional, there may be as many as six normal directions. 
Let $w(x^{\mu})$ be the six-vector specifying the amplitude of the 
oscillation at the spacetime point $x^{\mu}$, and let the oscillation 
itself be a superposition of $N$ waves:
$$w = \sum_{I=1}^{N} A_{I} \sin{ k_{I \mu} x^{\mu}} \  $$ 
(The component oscillations may be superposed because they are not 
assumed to be physical gravitational waves but are merely Fourier 
components used to build up the surface.) Then the 4-space metric  
is $$ g_{\mu \nu} = \gamma_{\mu \nu} + w,_{\mu}w,_{\nu} 
= \gamma_{\mu \nu} + \sum_{I,J} k_{I \mu} k_{J \nu} A_{I} 
A_{J} \cos{ k_{I \alpha} x^{\alpha}} \cos{ k_{J \beta} x^{\beta}}\ $$ 
where $\gamma_{\mu \nu}$ is the metric of the (flat) hyperplane tangent 
to the surface. It may be verified that $$ g^{\mu \nu} = 
(\gamma^{-1})^{\mu \nu} - H (\gamma^{-1})^{\mu \lambda} 
(\gamma^{-1})^{\nu \kappa} w,_{\lambda} w,_{\kappa} $$ 
where $$ H = 1 / (1 + (\gamma^{-1})^{\rho \sigma} w,_{\rho} w,_{\sigma} ).$$ 
In all of these equations, expressions such as $w_{I}w_{J}$ are to be 
interpreted as dot products in the normal part of hyperspace.

The salient point is that $w,_{\mu}w,_{\nu}$ is always a sum of the products 
of cosines. Therefore, if one averages over distances much larger than the 
longest wavelength $1/k_{I \mu},$ one has $$ <g_{\mu \nu}> = 
\gamma_{\mu \nu} + (1/2) \sum_{I=1}^{N} A_{I}^{2} k_{I \mu} k_{I \nu}\ ,$$ 
where the factor of one-half comes from the average value of the squared 
cosine.

Thus the elaborately dimpled surface looks flat in the large: hardly a 
surprise. However, an interesting effect can be produced if $\gamma_{\mu \nu}$
 contains negative terms: the waves can make the corresponding terms in 
$<g_{\mu \nu}>$ positive if the amplitude-to-wavelength ratios are large 
enough. In particular, if $\gamma_{\mu \nu}$ is the Minkowski metric, the 
oscillating terms can be chosen to make $<g_{\mu \nu}>$ Euclidean (or 
pseudo-Euclidean). If, on the other hand,  $\gamma_{\mu \nu}$ is 
pseudo-Euclidean, the oscillating terms can make $<g_{\mu \nu}>$ the 
metric of a Minkowski space with trace $\pm 2.$

It is instructive to look at the case of four waves of the same amplitude, 
each with $$ k_{I \mu} = k_{I} \delta_{I \mu}. $$ Then, $$ y = A( 
\sin{k_{1}x^{1}} + \sin{k_{2}x^{2}} + \sin{k_{3}x^{3}} + \sin{k_{4}x^{4}}) $$ 
and, $$ <g_{\mu \nu}> = \gamma_{\mu \nu} + (1/2)\sum_{I=1}^{N}A^{2}(k_{I})^{2}
 \delta_{I \mu} \delta_{I \nu}.$$
If  $\gamma_{\mu \nu}$ is the Minkowski metric with trace $+2,$ 
$<g_{\mu \nu}>$ will look Euclidean at large scales if $(k_{1})^{2} = 
(k_{2})^{2} = (k_{1})^{3} \equiv k^{2}$ and $$A^{2}(k_{4})^{2} = 4 + 
A^{2}k^{2}$$ In the same way, a pseudo-Euclidean metric will look 
Minkowskian at large scales if   $$A^{2}(k_{4})^{2} = 4 - A^{2}k^{2}.$$ 

It is natural to ask what sort of matter distribution would produce such an 
effect; actually, in the most trivial cases, no matter is required at all. 
For example, if $k = 0$ and $A = 2/k_{4},$ the metric is just   
$$ g_{\mu \nu} = \gamma_{\mu \nu} + (4/k_{4}^{2}) \cos{^{2}k_{4}x^{4}} 
\delta_{\mu 4} \delta_{\nu 4}.$$ 
This is simply a rescaling of the time coordinate if $\gamma_{\mu \nu}$ 
is diagonal: $$(d\bar{x}^{4})^{2} = (dx^{4})^{2}[(1 + 
(4/k_{4}^{2})\cos{^{2}k_{4}x^{4}}]/\gamma_{44}$$ The space has no 
curvature, much like a rippling flag.

A rather artificial example of a matter distribution in pseudo-Euclidean 
4-space which results in a Minkowskian geometry when averaged over large 
distances is the case $A^{2}k^{2} = 4$ and $ k_{4} = 0$. Then the time 
components of the metric have no effect, and: $$g_{ij} = - \delta_{ij} + 
4 \cos{kx^{r}} \cos{kx^{s}} \delta_{ir} \delta_{js} ; $$ $$ g^{ij} = - 
\delta^{ij} + 4 \cos{kx^{i}} \cos{kx^{j}}/ (-1 + 4 \cos{kx^{r}} 
\cos{kx^{s}} \delta_{rs} ).  $$The only non-vanishing Christoffel 
symbols are $$ \Gamma^{i}_{jj} = -J \sin{kx^{i}} \cos{kx^{j}}  $$ 
where $$ J = 4k/(-1 + 4(\cos{^{2}kx} + \cos{^{2}ky} + \cos{^{2}kz}  )).  
$$ The corresponding Ricci tensor has components:

$$ R_{xx} = J \sin{kx} [\sin{ky} + \sin{kz} - J(\sin{ky} \cos{^{2}ky} 
+ \sin{kz} \cos{^{2}kz } )]  $$ 

$$ R_{xy} = J^{2} \sin{kx} \cos{kx} \sin{ky} \cos{ky}   $$
plus three more found by cyclically permuting $x,$ $y,$ and $z$ in these 
expressions. 

It is hard to see how the resulting stress tensor could be established 
physically. More general oscillations producing curvature of the time 
coordinate would probably look even less physical, with the mass density 
also oscillating and taking on negative values. 

Nevertheless, transverse oscillations of spacetime may have some physical 
interest. Four interfering plane-wave like oscillations would create a 
cellular structure in spacetime on the scale of the wavelengths; this is 
very reminiscent of the various discrete-space approaches to quantum 
gravity, reviewed in [Gibbs 1995]. In addition, an undulating 
pseudo-Euclidean metric which looks Minkowskian in the large and which 
features alternating regions of positive and negative mass sounds remarkably 
like a geometrization of Winterberg's ``Planck aether'' {Winterberg 1994 
and 1995]. Finally, if rather obviously, transverse oscillations could be used
to Fourier-analyze embedded solutions of Einstein's equation.

\begin{center}
{\bf BIBLIOGRAPHY}
\begin{itemize}

\item P. A. M. Dirac 1975 {\it The General Theory of Relativity.} 
(Wiley-Interscience, New York)

\item P. E. Gibbs 1995 {\it hep-th/9506171}

\item D. Kramer {\it et al.} 1980 {\it Exact Solutions of Einstein's 
Field Equations.} (Cambridge University Press)

\item J. Rosen 1965 {\it Rev. Mod. Phys.,} 37(204) 

\item F. Winterberg 1994 {\it Int. J. Theor. Phys.,} 33(1273)

\item F. Winterberg 1995 {\it Int. J. Theor. Phys.,} 34(265)

\end{itemize}

\end{center}

\end{document}